\documentstyle [preprint,aps]{revtex}
\begin{document}
\draft
\title{HYDROTHERMAL SURFACE-WAVE INSTABILITY
AND THE KURAMOTO-SIVASHINSKY EQUATION}
\author{ R. A. Kraenkel, J. G. Pereira}
\address{Instituto de F\'{\i}sica Te\'orica\\
Universidade Estadual Paulista\\
Rua Pamplona 145\\
01405-900\, S\~ao Paulo\, SP --
Brazil}
\author{M. A. Manna}
\address{Laboratoire de Physique Math\'ematique\\
Universit\'e de Montpellier II\\
34095 Montpellier Cedex 05 --
France}
\maketitle
\begin{abstract}
We consider a system formed by an infinite viscous liquid layer with a constant
horizontal temperature gradient, and a basic nonlinear bulk velocity profile.
In the limit of long-wavelength and large nondimensional surface tension, we
show that hydrothermal surface-wave instabilities may give rise to disturbances
governed by the Kuramoto-Sivashinsky equation. A possible connection to
hot-wire experiments is also discussed.
\end{abstract}

\section{Introduction}

Thermocapillary dynamics in thin two-dimensional liquid layers, and in
particular the convective instabilities of such flows have been a subject of
much interest \cite{1,2}. When the upper free surface of a planar liquid,
bounded below by a rigid plate and above by an interface with a passive gas, is
submitted to a temperature gradient, a corresponding gradient in the surface
tension will appear which will produce  motion in the bulk fluid. If a
vertical temperature gradient is applied to a basic static state, convective
motion sets in, a phenomenon called Marangoni convection. However, a horizontal
temperature gradient may also give rise to instabilities, provided the basic
state is not static \cite{1}. These are the so called hydrothermal
instabilities, which are a coupled effect produced by both temperature
and velocity gradients. Smith and Davis \cite{1,2} identified two such
instabilities, which manifest themselves in the form of convection
and surface-wave motion.

Our concern in this letter will be the study of the hydrothermal surface-wave
instability. It has been shown in Ref.\cite{1} that it is possible to have
such instabilities characterized by a zero wave-number, provided
the basic underlying flow is a nonlinear return flow.
The long wavelenght nature of the instability
allows us to broach the problem by a long-wave perturbative analisys, where
nonlinearity may be taken into account. By using the reductive perturbation
method of Taniuti \cite{3}, we will show that, in the limit of large
nondimensional surface tension, the waves originated by this instability turn
out to be governed by the Kuramoto-Sivashinsky equation \cite{4,5}.
This equation has
already been derived in different physical contexts \cite{6}. In particular it
has been obtained as the equation governing perturbations from a reference
Poiseuille flow of a film layer on an inclined plane \cite{7}, also in the
specific limit of large nondimensional surface tension \cite{8}. However, in
this case, the instability under consideration was hydrodynamical in nature,
and not hydrothermal, as is the case considered in the present work.

The basic interest of the results presented in this work is concerned with
recent
experimental evidences for the existence of surface hydrothermal waves. For
example, it has been reported in the literature the observation of such waves
in
a cylindrical container \cite{9}. Moreover, hot-wire experiments performed
recently \cite{10,11} have indicated the presence of propagative patterns. In
the
final section, we will speculate on a possible connection between our
results and these experiments.
\section{Mathematical Formulation}

We consider a fluid layer of height $d$, bounded below, at $z=0$, by a rigid
perfectly insulating plate, and above, at $y = d + \eta(x,t)$, by a free
deformable surface in contact with a passive gas of negligible density and
viscosity. The liquid is characterized by a density $\rho$, thermal
conductivity $k$, thermal diffusivity $\kappa$, unit thermal surface
conductance $h$, and dynamic viscosity $\mu$. To the free surface we
associate a surface tension $T$, which will be assumed to depend
on the local temperature $\theta$ according to the linear law
\begin{equation}
T = T_0 - \gamma \left(\theta - \theta_0 \right),
\end{equation}
where $\gamma$ is a positive constant, and $T_0$, $\theta_0$ are reference
values for surface tension and temperature, respectively.

We will be concerned with effects coming from thermocapillarity only. Therefore
we will neglect gravity. This is a good approximation for a thin enough layer,
or a layer in a microgravity environment. The equations governing the fluid
motion are written as:
\begin{equation}
u_x + w_z = 0 \, ,
\end{equation}
\begin{equation}
\rho \left(u_t + u u_x + w u_z \right) = - p_x + \mu \left(u_{xx} + u_{zz}
\right) \, ,
\end{equation}
\begin{equation}
\rho \left(w_t +u w_x + w w_z \right) = - p_z + \mu \left(w_{xx} + w_{zz}
\right) \, ,
\end{equation}
\begin{equation}
\theta_t + u \theta_x + w \theta_z = \kappa \left(\theta_{xx} + \theta_{zz}
\right) \, ,
\end{equation}
where the subscripts denote partial differentiation, $u$ and $w$ are,
respectively, the $x$ and $z$ components of the velocity, and $p$ is the
pressure. On the upper free surface, these equations are
subject to the following boundary conditions
\begin{equation}
\eta_t + u \eta_x = w \, ,
\end{equation}
\begin{equation}
p - {2 \mu \over {N^2}} \left[w_z + u_x \left(\eta_x \right)^2 - \eta_x
\left(u_z + w_x \right) \right] = - {T \over{N^3}} \eta_{xx} \, ,
\end{equation}
\begin{equation}
\mu \left[1 - \left(\eta_x \right)^2 \right] \left[u_z + w_x \right] + 2 \mu
\eta_x \left(w_z + u_x \right) = N \left[T_x + \eta_x T_z \right] \, ,
\end{equation}
\begin{equation}
\eta_x \theta_x + \theta_z = {h \over k} \left( \theta - \theta_{\infty}
\right),
\end{equation}
with
$$ N = \left[1+ \left( \eta_x \right)^2 \right]^{1\over2} \, .$$
Equation (6) is the kinematic boundary condition, Eqs. (7) and (8) represent
the normal and tangential stress balance at the upper free surface, and Eq. (9)
is the equality of the heat flux at the interface,
where $\theta_{\infty}$ is the temperature at $z \rightarrow \infty$.
On the lower plate, the boundary conditions are
\begin{equation}
u = w = \theta_z = 0 \, ,
\end{equation}
which characterizes the no slip and zero heat flux conditions.

Perturbation theory will be performed on a basic state of the system, defined
by imposing to the infinite liquid layer a constant horizontal temperature
gradient
\begin{equation}
{d \theta_b \over d x} = - \beta \, ,
\end{equation}
with $\beta$ a constant, which will give rise to a basic velocity profile in
the
bulk fluid. Moreover, we will assume a nonlinear velocity profile since, as
remarked in Ref. \cite{1}, a linear flow profile is not susceptible
to long-wave instabilities.

We now proceed to a nondimensionalization of the equations and boundary
conditions. To this end, we take $d$ as the unit of lenght, $\mu/
\gamma \beta $ as the unit of time,
$\beta d$ as the unit of temperature, $\gamma \beta$ as the unit of
pressure, and $T_0$ as the unit of surface tension. In this process the
following dimensionless numbers appear:
$$
R = {\rho \gamma \beta d^2 \over \mu^2} \, , \quad \sigma =
{\mu \over \rho \kappa} \, , \quad B = {h d \over k} \,, \quad
S = {\rho d T_0 \over \mu^2} \, ,
$$
where $R$ is the Reynolds number, $\sigma$ is the Prandtl number, $B$ is the
Biot number, and $S$ is the surface tension number. In addition, we have also
the Marangoni number $M$, which is defined by $M=R\sigma$.

The basic state, from which we will consider small disturbances, is the return
flow solution to the Eqs.(2-10) which, in the limit $S \rightarrow \infty$,
is given by:
\begin{mathletters}
\begin{equation}
u_b = {3 \over 4} z^2 - {1 \over 2} z
\end{equation}
\begin{equation}
w_b = 0 \,; \quad p_b = {3 \over 2} x
\end{equation}
\begin{equation}
\theta_b - \theta_0 = -x + R \sigma \left[{1 \over 16} \left(1 - z^4 \right) -
{1 \over 12} \left(1 - z^3 \right) \right]
\end{equation}
\begin{equation}
\theta_{\infty} = - x \,; \quad \eta_b = 0 \,.
\end{equation}
\end{mathletters}

This solution has been obtained in Ref.\cite{12}. It is exact in the limit of
infinite surface-tension. However, in what follows we will consider $S$ to be
large, albeit finite. Accordingly, we should introduce corrections to Eqs.(12)
up to some appropriate order of $S^{-1}$. These corrections, though, are of
higher orders than those contributing to our calculations, except in Eq.(7). In
this equation, the right hand side is proportional to $S$. To match orders
correctly we make use of the basic solution $\eta_b$ correct up to order
$S^{-1}$, which is given by \cite{12}
$$\eta_b = - {R \over 4 S} x^3 + {R \over 16 S} x \,.$$
It is important, at this point, to note that the range of validity
of the basic solution is not arbitrary. As shown in Ref. \cite{12}, the
largeness of $S$ is connected to the largeness of the aspect ratio A, the
ratio of the width of the layer to its depth. In fact, we must have $S
\sim {\cal O}(A{^4})$ as $A \rightarrow \infty$, so that we are not allowed
to take $ x \rightarrow \infty$ independently of $S$.

\section{Perturbation Theory}

Let us consider the following perturbations to the basic state :
\begin{mathletters}
\begin{equation}
u = u_b + \epsilon \left(u_0 + \epsilon u_1 + .... \right) \,,
\end{equation}
\begin{equation}
w = \epsilon^2 \left(w_0 + \epsilon w_1 + ...\right) \,,
\end{equation}
\begin{equation}
p = p_b + \epsilon \left(p_0 + \epsilon p_1 + ...\right) \,,
\end{equation}
\begin{equation}
\theta = \theta_b + \epsilon \left(\theta_0 + \epsilon \theta_1 +...\right)
\,,
\end{equation}
\begin{equation}
\eta = \eta_b + \epsilon \left(\eta_0 + \epsilon \eta_1 + ... \right) \,,
\end{equation}
\end{mathletters}
where $\epsilon$ is a small parameter. Next, we introduce the slow variables
according to
\begin{equation}
\xi = \epsilon \left(x - ct \right) \,,
\end{equation}
\begin{equation}
\tau = {\epsilon^2} t \,,
\end{equation}
and we suppose that the perturbations depend on $(x,t)$ through $ (\xi,\tau)$
only. We now make more precise the meaning of {\it large albeit finite} for
$S$, by assuming that
\begin{equation}
S = \epsilon^{-2} \overline{S} \,,
\end{equation}
with $\overline{S} \sim {\cal O}(\epsilon^0)$.

We are able, at this point, to obtain an order by order solution to the
problem by integrations in the $z$ variable. At each order, the bulk equations
and the boundary conditions at the bottom yield $u, w, \theta, p$ in
terms of three arbitrary functions, which will be determined in
terms of $\eta_0$ by
the boundary conditions at the upper free surface. However, as we have four
equations, a compatibility condition will arise at each order. Explicitly,
at order $\epsilon^0$, we get:
\begin{mathletters}
\begin{equation}
u_0 = -{3\over 2} z \eta_0 \,;
\end{equation}
\begin{equation}
w_0 = {3\over 4} z^2 \eta_{0 \xi} \,;
\end{equation}
\begin{equation}
\theta_0 = R \sigma \left[{z^3 \over 4} + \left({1\over 2B} - {1\over 4}
\right) \right] \eta_0 \,;
\end{equation}
\begin{equation}
p_0 = - {{\overline S} \over R} \eta_{0 \xi \xi} \,.
\end{equation}
\end{mathletters}
{}From Eqs. (6) and (8), we get as the compatibility condition at this order
\begin{equation}
c = -{1\over 2} \,.
\end{equation}

At the next order, the relevant solutions are
\begin{mathletters}
\begin{equation}
u_1 = {R\over 32}\left(z^4 - 4 z^3 \right) \eta_{0 \xi} - {{\overline S}
\over 2R} z^2 \eta_{0 \xi \xi \xi} + z f \,,
\end{equation}
\begin{equation}
w_1 = -{R \over 160} \left(z^5 - 5 z^4\right)\eta_{0\xi\xi} + {{\overline S}
\over 6R}z^3 \eta_{0\xi\xi\xi\xi} - {z^2 \over 2} f_\xi \,,
\end{equation}
\end{mathletters}
with $f = f(\xi,\tau)$ an arbitrary function. Also at this order, from Eq.(6),
we have :
\begin{equation}
{3 \over 4}\eta_{1 \xi} + {1\over 2} f_\xi = - \eta_{0\tau} + {5\over 4}
\eta_0 \eta_{0\xi} + {R\over 40}\eta_{0\xi\xi} + {{\overline S} \over 6R}
\eta_{0\xi\xi\xi\xi}.
\label{aux}
\end{equation}
{}From Eq.(8), we get:
\begin{equation}
{3\over 4}\eta_{1 \xi} + {1\over 2} f_\xi = {R\over 4}\left({\sigma \over B} +
{1\over 2}\right)\eta_{0\xi\xi} + {{\overline S}\over
2R}\eta_{0\xi\xi\xi\xi}\,.
\end{equation}
Substituting into Eq.(\ref{aux}), we obtain an evolution equation for $\eta_0$:
\begin{equation}
\eta_{0 \tau} - {5\over 4}\eta_0 \eta_{0\xi} + R\left({\sigma \over 4B} +
{1\over 10}\right)\eta_{0\xi \xi} + {{\overline S}\over 3R}\eta_{0\xi\xi\xi\xi}
=0 \,.
\label{kura1}
\end{equation}
Returning to the laboratory coordinates $(x,t)$, and remembering that
${\overline S} = {\epsilon^2} S$ and $\eta_0 = \epsilon^{-1} \eta$,
Eq.(\ref{kura1})
reads
\begin{equation}
\eta_t + \left(c - {5\over 4}\eta \right)\eta_x + R\left({\sigma \over 4B}
+{1\over 10}\right)\eta_{xx} + {S\over 3R}\eta_{xxxx} = 0 \,.
\label{kura2}
\end{equation}
Defining a new field by
$$ \zeta = c - {5 \over 4}\eta \,, $$
Eq.(\ref{kura2}) becomes :
\begin{equation}
\zeta_t + \zeta \zeta_x + R\left({\sigma \over 4B} + {1\over 10}\right)
\zeta_{xx} + {S \over 3R}\zeta_{xxxx} = 0 \,.
\label{kura3}
\end{equation}
By a simple rescaling of variables, this equation can be written as
\begin{equation}
\zeta_t + \zeta \zeta_x + \zeta_{xx} + \zeta_{xxxx} = 0 \,,
\end{equation}
which is the Kuramoto-Sivashinsky (KS) equation in its standard form
\cite{4,5}.

The linear analysis performed by Smith and Davis \cite{1} can be recovered from
Eq.(24) if we neglect the nonlinear term, and take
$$ \zeta \sim exp \left[\imath \left(\alpha x - \omega t \right)
\right] \,. $$
This leads to the dispersion relation
\begin{equation}
\omega = \imath \left[\alpha^2 R\left({\sigma \over{4B}} + {1 \over 10}
\right) - \alpha^4 {S\over 3R}\right] \,,
\end{equation}
from which we can see that the second derivative and the fourth derivative
terms of Eq.(\ref{kura3}) are, respectively, antidissipative and dissipative.
Therefore
there exists a critical point at the long wavelenght limit, given by $ \omega =
0$, in which these terms compensate each other. In this limit,
$\alpha \rightarrow 0$, and assuming that $S = \alpha^{-2}{\overline S}$,
we obtain the critical Reynolds number
\begin{equation}
R_c = \left[{{\overline S} \over 3} \left({\sigma \over 4B} + {1\over 10}
\right)^{- 1} \right]^{1\over2} \,,
\end{equation}
which is the result of Smith and Davis \cite{1}.

\section{Discussion}

The KS equation is a non-integrable equation, despite the fact that some
exact solutions have been found \cite{4,13}. Generically, the solutions
to the KS equation are characterized by the coexistence of
coherent spatial structures with temporal chaos \cite{14}. In particular,
the search for a better understanding of
the long wavelength properties of the KS equation has been intensively pursued
lately \cite{15}. And it is precisely in the long-wave limit that we have
obtained  the KS equation as governing hydrothermal surface-wave
instabilities in a thin viscous layer of fluid with  horizontal temperature and
velocity gradients, and in the limit of large nondimensional surface-tension.
This result could possibly be connected to a recent
experiment \cite{9} which has been performed to study the features of
hydrothermal waves appearing when a cylindrical liquid layer is laterally
heated. The existence of hydrothermal wave instabilities has
been observed experimentally,
and the results were found to partially agree with the previous
theoretical predictions of Smith and Davis \cite{1}.
If we believe that, despite the
difference in the geometry the results presented in Ref.\cite{9} can somehow be
considered as an approximation to the theoretical predictions, then the
long-wave dynamics
of the free surface described by the KS equation should play a role in the
explanation of the propagative patterns observed in this kind of experiment.

At last, we would like to speculate on a possible connection between the
results obtained in this letter and hot-wire experiments \cite{10,11}.
These experiments
consist of a hot wire placed horizontally just below a free surface of a
fluid. Above a certain
critical value of the electrical power supplied to the hot wire, propagating
patterns are observed. The existence of both horizontal temperature gradients
and a
basic velocity flow, as the fluid is convecting, makes contact with the
theoretical setting in which hydrothermal waves show up. Of course, this is not
enough to ensure that the above results can have an immediate application
to the hot-wire experiments. However, since the system under consideration here
presents similarities with the above mentioned experiments, they could
help to elucidate the physical mechanism responsible for the
observed propagative patterns. The idea to mimic
features of a convecting fluid by a two-dimensional system with a velocity
gradient has been examined by Roz\`{e} \cite{16}, based on data of an
experiment with a much shorter wire \cite{17}. However, this approach,
which used a linear flow profile and linearized equations, turned out
to be rather disappointing. In fact, as shown in Ref.\cite{1}, there is no
long-wave instability in this case. Only when a nonlinear return flow
solution is supposed,
which is indeed more appropriate for a closed system, long-wave hydrothermal
phenomena appear. Therefore, we think it would be interesting to reexamine
the ideas proposed in Ref.\cite{16} for the hot wire problem in the light of
the present results.

In summary, we have made in this letter a nonlinear analysis of the long
wavelength hydrothermal instabilities, which can be considered as an
extension of a problem first discussed by Smith and Davis
\cite{1} in a linear approximation. We have shown that the free-surface
dynamics is governed by the Kuramoto-Sivashinsky equation, and that the
previous results can be recovered from ours as a particular case.

\acknowledgements

Two of the authors (R.A.K. and J.G.P.)  would like to thank Conselho Nacional
de
Desenvolvimento Cient\'{\i}fico e Tecnol\'ogico (CNPq), Brazil, for
partial support.

\end{document}